\documentclass[sigconf]{acmart}
\usepackage{cleveref}

\usepackage{enumitem}
\usepackage{mleftright}
\usepackage{placeins}
\usepackage{tcolorbox}
\tcbuselibrary{listings}

\renewcommand\footnotetextcopyrightpermission[1]{} 
\setcopyright{acmlicensed}
\copyrightyear{2018}
\acmYear{2018}
\acmDOI{XXXXXXX.XXXXXXX}
\acmConference[Conference acronym 'XX]{Make sure to enter the correct
  conference title from your rights confirmation email}{June 03--05,
  2018}{Woodstock, NY}
\acmISBN{978-1-4503-XXXX-X/18/06}

\begin{document}
\title{Analytical Search}

\author{Yiteng Tu}
\affiliation{%
    \institution{DCST, Tsinghua University}
    \institution{Quan Cheng Laboratory}
    \city{Beijing}
    \country{China}
}
\email{tyt24@mails.tsinghua.edu.cn}

\author{Shuo Miao}
\affiliation{%
    \institution{DCST, Tsinghua University}
    \city{Beijing}
    \country{China}
}

\author{Weihang Su}
\affiliation{%
    \institution{DCST, Tsinghua University}
    \city{Beijing}
    \country{China}
}

\author{Yiqun Liu}
\affiliation{%
    \institution{DCST, Tsinghua University}
    \city{Beijing}
    \country{China}
}

\author{Qingyao Ai*}
\affiliation{%
    \institution{Quan Cheng Laboratory}
    \institution{DCST, Tsinghua University}
    \city{Beijing}
    \country{China}
}
\email{aiqy@tsinghua.edu.cn}

\begin{abstract}
Analytical information needs, such as trend analysis and causal impact assessment, are prevalent across various domains including law, finance, science, and much more. 
However, existing information retrieval paradigms, whether based on relevance-oriented document ranking or retrieval-augmented generation (RAG) with large language models (LLMs), often struggle to meet the end-to-end requirements of such tasks at the corpus scale.
They either emphasize information finding rather than end-to-end problem solving, or simply treat everything as naive question answering, offering limited control over reasoning, evidence usage, and verifiability. 
As a result, they struggle to support analytical queries that have diverse utility concepts and high accountability requirements.

In this paper, we propose \textbf{analytical search} as a distinct and emerging search paradigm designed to fulfill these analytical information needs. 
Analytical search reframes search as an evidence-governed, process-oriented analytical workflow that explicitly models analytical intent, retrieves evidence for fusion, and produces verifiable conclusions through structured, multi-step inference. 
We position analytical search in contrast to existing paradigms, and present a unified system framework that integrates query understanding, recall-oriented retrieval, reasoning-aware fusion, and adaptive verification. 
We also discuss potential research directions for the construction of analytical search engines. 
In this way, we highlight the conceptual significance and practical importance of analytical search and call on efforts toward the next generation of search engines that support analytical information needs.
\end{abstract}



\maketitle

\section{Introduction}
Information Retrieval (IR) has long been centered on a fundamental goal: assisting users in accomplishing tasks by organizing, retrieving, and utilizing information to satisfy their needs~\cite{manning2008introduction,shah2025ta,taylor1962process,ai2023information}. 
These needs span a broad spectrum, ranging from simple fact lookup, complex exploratory and decision making process~\cite{marchionini2006exploratory,singer2013ordinary}. 
Classical IR systems based on relevance ranking have been proven to be highly effective for ad-hoc retrieval tasks in which the user seeks documents that are topically relevant to a query~\cite{robertson2009probabilistic,karpukhin2020dense,ponte2017language,zhan2020repbert}. 
Even with the advent of conversational search~\cite{radlinski2017theoretical,mo2025survey} and session search~\cite{chen2022overview,chen2019tiangong} systems where models can exploit the user feedback or context information, the dominant interaction paradigm remains largely unchanged: systems return a ranked list of results, and users manually inspect, synthesize, and reason over the retrieved information to complete their tasks~\cite{liu2021deconstructing}. 
This paradigm is well crafted to support naive \textit{information finding}, where information needs can be solved by examining one or a few relevant documents that contain the answers to the user's questions. 

However, real-world information needs of users often go beyond naive information finding~\cite{metzler2021rethinking,singer2013ordinary,tang2022re,ai2023information}. 
In domains such as law, finance, scientific research, and political analysis, users often pose analytical, exploratory, and decision-oriented questions that require synthesizing evidence across sources, performing comparisons or aggregations, and drawing reasoned conclusions~\cite{luo2025towards,hu2025drama}. 
For instance, queries such as "\textit{What was the total number of theft incidents reported on public transit over the past year?}" or "\textit{How did News A influence Stock B?}" may not be able to find answers from one or a few documents, but requires collecting heterogeneous evidence, aligning it along temporal or causal dimensions, and applying multi-step reasoning to synthesize the final answer.
We refer to these types of needs as \textbf{analytical information needs}.
Historically, to solve analytical problems, users must rely on a labor-intensive workflow: issuing multiple search queries against search engines or databases, manually filtering and validating retrieved materials, and incrementally assembling evidence to support an analysis. 
This process is not only time-consuming and cognitively demanding, but also difficult to scale and reproduce.

Fortunately, recent advances in large language models (LLMs)~\cite{touvron2023llama,bai2023qwen,achiam2023gpt,hurst2024gpt,guo2025deepseek,yang2025qwen3} have provided new opportunities for analytical information needs. 
The integration of LLMs with IR systems, through paradigms such as retrieval-augmented generation (RAG)~\cite{borgeaud2022improving,su2024dragin,lewis2020retrieval,guu2020retrieval,izacard2021leveraging,su2025parametric}, search agents~\cite{li2025search,jin2025search,li2025webthinker,chen2025learning}, and tool-augmented reasoning frameworks~\cite{ma2024sciagent,li2023api,parisi2022talm}, has substantially expanded the scope of tasks that modern IR systems can do. 
Similar to humans, LLMs can understand both the deep semantics of text and filter documents based on the needs of input queries with complex reasoning processes.
They have the potential to largely replace human efforts in analytical tasks, fully automate the labor-intensive workflow that used to be necessary, and shift IR systems from an information finding tool to a true problem-solving system.

Nevertheless, simply adapting LLMs or coupling them with retrieval components is insufficient for solving analytical information needs in practice. 
Theoretically, analytical information needs could be solved by using a strong LLM to examine all documents in a corpus and synthesize results based on the needs of user queries. 
Empirically, such a naive method is not feasible considering the extreme computational cost of LLMs and the large scale of retrieval collections.
In our preliminary experiment, employing Qwen-32B~\cite{bai2023qwen,yang2025qwen3} to identify instances of voluntary surrender in ten thousand legal case documents requires over 3 hours on 2 A100 GPUs. 
This inference cost is prohibitive for scaling to a corpus of millions of documents.
While equipping LLMs with existing search tools to conduct RAG~\cite{lewis2020retrieval,guu2020retrieval,izacard2021leveraging} seems to be a straightforward solution, many studies have shown that existing RAG paradigms are incapable in such complex analytical tasks~\cite{luo2025towards,hu2025drama,sun2025agenticdata}.
Their performance is often limited by the small set of "top" documents, which are retrieved by models optimized for topical relevance but not analytical value.
More importantly, analytical information needs often involve document filtering and analysis from multiple aspects that require different types of querying and retrieval systems. 
Existing RAG paradigms intentionally decouple the retrieval process from the generation process, where LLMs don’t know how to effectively use multiple types of retrieval tools, and retrieval systems don’t know how to construct indexes and optimize the efficiency of the whole analytical workflow. 

In this paper, we argue that a new generation of search techniques is needed for analytical information needs, which we refer to as \textbf{analytical search}. 
Analytical search moves beyond naive relevance-oriented document retrieval~\cite{karpukhin2020dense,robertson2009probabilistic} toward systematic frameworks in which search is understood as a process of assembling the whole picture rather than finding an isolated puzzle piece.
This means rethinking search engine design from multiple perspectives, involving but not limited to: 
(1) Analytical query processing, how to decompose analytical queries to actionable items, utilize multiple types of retrieval tools, and adjust plans iteratively based on preliminary results and feedback; 
(2) Evidence-oriented retrieval pipeline, how to optimize recall in a multi-path retrieval workflow, integrate the retrieved evidence while providing traceable and auditable process signals, and conduct efficient evidence verification with retrieval and generation models jointly; 
and (3) Dynamic and reasoning-enhanced indexing, how to cache the reasoning process of data analysis, adapt the index structure to online request distributions, and continually improve both the effectiveness and efficiency of analytical search. 
To better illustrate the unique importance and challenges of analytical search, we discuss and differentiate it from related concepts such as RAG, deep research, agentic databases, etc., and present an example system framework that captures the core components and interactions required to support analytical search. 
We outline open research challenges and future directions, and hope to attract more community efforts to look into this critical and relatively underexplored area of information retrieval.

\section{Scope, Objectives, and Positioning} \label{sec:what}
\subsection{Analytical Information Needs}
A defining characteristic of analytical search lies in the nature of the information needs it seeks to address.
\textbf{Analytical information needs} refer to user queries that require systematic reasoning over multiple pieces of evidence, including the explicit modeling of implicit intent, the decomposition of complex problems, and the synthesis and validation of conclusions rather than the simple retrieval of isolated facts.
They manifest in several recurring forms that differ in their analytical depth, reasoning requirements, and decision implications. 
Following the taxonomy widely used in data analytics and statistics~\cite{el2019descriptive,lepenioti2020prescriptive,sharma2022analytics}, we broadly group them into: 
\begin{itemize}[leftmargin=*]
    \item \textbf{Descriptive Analytical Needs}.
    Descriptive analytics focuses on understanding and summarizing what has happened, often through aggregation, comparison, and pattern identification. 
    It aims to construct an interpretable representation of historical data or evidence, frequently serving as the foundation for deeper analysis.
    Descriptive queries may involve simple operations such as counting, ranking, or filtering, but often extend to more complex descriptive analyses, including temporal trends and multi-faceted comparisons. 
    Even when the underlying operations are conceptually simple, there are still many challenges, such as identifying appropriate data sources, aligning definitions, and ensuring evidential consistency. 
    Examples:
    \newline - \textit{How often did people get robbed on the bus or subway last year?}
    \newline - \textit{Which cities experienced the highest growth in public transportation crime rates?}
    \newline - \textit{What are the most frequently cited reasons for product returns across different regions?}
    \newline
    Instead of finding single facts, these queries often involve analysis like structured aggregation across time, entities, and sources.

    \item \textbf{Predictive Analytical Needs}.
    Predictive analytics aims to predict future outcomes or latent relationships based on observed evidence. 
    Predictive queries require the system to move from summarization to future inference. 
    This often involves causal reasoning, correlation analysis, or extrapolation from historical patterns.
    Predictive analytical queries typically require identifying relevant signals, assessing their reliability, and integrating evidence across heterogeneous sources. 
    While full statistical modeling may not always be needed, predictive analytical search needs to reason about uncertainty, temporal dependency, and potential confounding factors.
    Examples:
    \newline - \textit{Did the introduction of traffic cameras lead to a reduction in pedestrian accidents?}
    \newline - \textit{How is News A likely to affect Stock B in the short term?}
    \newline - \textit{Based on historical data, is the current trend in influenza cases likely to continue?}
    \newline
    These queries naturally incorporate causal analysis or correlation-based reasoning, seek not only to describe surface phenomena, but to infer underlying relationships and anticipate future states.
    
    \item \textbf{Prescriptive Analytical Needs}.
    Prescriptive analytics represents the most decision-oriented form of analytical reasoning. 
    Queries in this category aim to evaluate alternative actions, policies, or strategies, and to provide guidance on what should be done under specific constraints or objectives. 
    In analytical search, prescriptive queries often build upon descriptive and predictive analyses, requiring integrated reasoning across evidence, outcomes, and trade-offs.
    Such queries are especially prevalent in domains such as policy analysis, business strategy, and operational decision-making. 
    They demand not only accurate evidence retrieval and inference, but also explicit modeling of objectives, constraints, and evaluation criteria.
    Examples:
    \newline - \textit{Given recent market conditions, should an investor increase exposure to Stock C?}
    \newline - \textit{Which policy intervention would most effectively reduce theft incidents on public transportation?}
    \newline - \textit{Which COVID-19 mitigation strategy would best balance public health outcomes and economic activity?}
    \newline
    These queries require generating analytical insights to support actionable conclusions, and the answers must be inherently contingent, context-dependent, and justified through transparent reasoning over evidence.
    
\end{itemize}

Together, these categories illustrate the breadth and complexity of analytical information needs. 
More importantly, they expose a set of fundamental challenges that arise when attempting to resolve such needs in an automated or semi-automated manner. 
Unlike conventional information-seeking tasks, the difficulty of analytical information needs does not stem from any single component, but from the compounded challenges distributed across the entire analytical workflow—from query understanding to evidence retrieval, fusion, and validation:
\begin{itemize}[leftmargin=*]
\item \textbf{Implicit and Complex Analytical Intent}.
Analytical information needs are frequently expressed in natural language that underspecifies critical analytical assumptions, constraints, and objectives. 
Users often omit temporal scopes, comparison baselines, evaluation criteria, or even the precise analytical goal. 
For example, a query such as "Has policy A improved urban air quality?" implicitly assumes a time window, a definition of improvement, and one or more baseline conditions. 
This creates a fundamental challenge at the query understanding stage: systems must reconstruct latent analytical intent rather than merely interpret surface-level query semantics. 
Failure to do so leads to incomplete or misaligned downstream analysis.
On the other hand, analytical queries rarely correspond to a single retrievable fact and require decomposition into multiple interdependent sub-questions, such as identifying relevant entities, collecting longitudinal evidence, performing comparisons, or estimating effects.
These sub-tasks are not independent: errors or omissions in early decomposition can propagate through the analytical pipeline. 
Designing systems that can reliably decompose complex analytical needs into executable and logically coherent sub-tasks also remains a core challenge.

\item \textbf{Beyond Naive Ad-hoc Retrieval}.
At the evidence collecting stage, analytical information needs impose substantially higher demands than traditional ad-hoc search. 
Potentially useful evidence is often sparse, distributed across heterogeneous sources, and may be weakly aligned with the original query at the surface level.
Critical evidence may reside in structured databases, statistical reports, policy documents, or unstructured narratives, each requiring different retrieval strategies. 
Moreover, analytically important evidence may not be topically prominent, making precision-oriented top-$k$ retrieval insufficient. 
Ensuring sufficient recall of analytically critical evidence, while managing noise and scale, constitutes a central retrieval challenge.

\item \textbf{Reasoning-intensive Fusion}.
Even when relevant evidence is successfully collected, it still requires fusion strategies that go beyond simple aggregation or summarization. 
People or systems must perform multi-step reasoning operations such as filtering, temporal alignment, comparison across alternatives, causal attribution, trend extrapolation, or trade-off analysis. 
These reasoning-based fusion steps often depend on intermediate representations and partial results, sometimes with the aid of external tools, rather than a single pass over retrieved content. 
Designing fusion mechanisms that are robust to incomplete or partially conflicting evidence is also a major challenge.

\item \textbf{Rigorous Conclusion Verification}.
Analytical conclusions are rarely self-evident from individual pieces of evidence. 
Instead, they must be validated through consistency checking, cross-source corroboration, or even sensitivity analysis. 
Conflicting evidence, data quality issues, or ambiguous causal signals are common in real-world analytical tasks. 
As a result, it imposes strong requirements on verification: systems must not only generate conclusions, but also assess whether those conclusions are sufficiently supported, identify potential weaknesses, and, when necessary, trigger additional retrieval or revision.
\end{itemize}

\subsection{Analytical Search}
Analytical search is the search paradigm that aims to resolve analytical information needs by explicitly modeling reasoning demands, retrieving and organizing evidence for analysis, and producing verifiable conclusions through structured, multi-step problem-solving.
Unlike traditional information retrieval systems that prioritize topical relevance and ranked document lists, analytical search treats search as an end-to-end analytical process. 
Its objective is not merely to surface potentially relevant information, but to support users in constructing justified conclusions through evidence-backed reasoning. 
This shift gives rise to several characteristics:

\begin{itemize}[leftmargin=*]
\item \textbf{Conclusion-oriented}.
The primary goal of analytical search is not to retrieve isolated facts, but to generate conclusions that synthesize multiple pieces of evidence. 
Accordingly, an analytical system is explicitly optimized for the correctness, completeness, and justification of the conclusion, rather than the linguistic fluency or stylistic quality of the response, unlike traditional QA systems. 
While natural language generation remains an important interface component, it is subordinated to the quality of the underlying analysis. 
A concise, well-supported conclusion is preferred over a verbose but weakly grounded answer. 
This orientation reflects a shift from answer-centric evaluation toward outcome-centric assessment, where the validity of conclusions and the soundness of the supporting reasoning take precedence.

\item \textbf{Complex Relevance}.
In analytical search, relevance is no longer determined primarily by surface-level lexical or semantic similarity between queries and documents. 
Instead, relevance is defined by a piece of information’s utility for reasoning and analysis. 
A document may be topically related yet analytically irrelevant if it does not contribute to answering a sub-question, supporting a hypothesis, or constraining an inference. 
Conversely, evidence that appears only weakly related at the surface level may be crucial for completing a reasoning chain. 
This notion of complex relevance requires retrieval systems to assess documents in terms of their logical role, evidential value, and compatibility with downstream reasoning processes, fundamentally extending the traditional IR relevance model.

\item \textbf{Evidence-governed}.
Finally, in analytical search, conclusions produced by the system must be grounded in verifiable, traceable evidence drawn from trusted data sources. 
Unlike generative search systems that may prioritize fluent answer generation, analytical search emphasizes evidential accountability: each claim, comparison, or quantitative result should be supported by explicit evidence that can be inspected and validated. 
This property is particularly critical in high-stakes domains such as policy analysis, finance, law, and scientific inquiry, where unsupported or hallucinated outputs are unacceptable.
\end{itemize}

\subsection{Differences with Existing Search Paradigms}
Analytical search does not emerge in isolation; rather, it intersects with several recent paradigms that combine retrieval, fusion, and generation.
However, despite superficial similarities, analytical search is conceptually and operationally distinct from existing approaches. 
Here, we clarify these distinctions by contrasting analytical search with three closely related paradigms. 

\subsubsection{Analytical Search vs. Retrieval-Augmented Generation (RAG)} \hfill \break
Retrieval-Augmented Generation (RAG)~\cite{borgeaud2022improving,su2024dragin,lewis2020retrieval,guu2020retrieval,izacard2021leveraging,su2025parametric} was originally proposed to improve the factual grounding of large language model outputs by conditioning generation on retrieved documents~\cite{tu2025rbft,su2024unsupervised}. 
In RAG-style systems, retrieval primarily serves as supporting context for answer generation, mitigating hallucinations and enhancing factual correctness. 
The dominant optimization target is the quality of the generated answer—typically measured by fluency, relevance, or factual accuracy at the surface level.

Analytical search, on the other hand, departs from this framing in several fundamental ways. 
It aims at optimizing problem-solving correctness and traceability, prioritizing whether the final conclusion is logically sound, evidentially supported, and reproducible.
Rather than treating retrieval as auxiliary input to a generator, analytical search treats retrieval as evidence construction. 
Besides, retrieved evidence is not merely prompts for generation but is explicitly selected, organized, and validated to support downstream reasoning. 
It is expected to reason over retrieved evidence through structured, multi-step processes such as decomposition, comparison, aggregation, and validation.
In this paradigm, reasoning is a first-class objective, not an implicit by-product of generation. 
As a result, analytical search systems must explicitly model reasoning demand and evidential roles—capabilities that are largely outside the design scope of conventional RAG pipelines.

\subsubsection{Analytical Search vs. Deep Research} \hfill \break
Deep research~\cite{huang2025deep,zhang2025deep,li2025webthinker} systems are designed to support open-ended exploration and knowledge accumulation, often aiming to produce comprehensive reports, background summaries, or literature-style syntheses. 
Their primary objective is breadth and coverage: gathering diverse perspectives, accumulating information across sources, and presenting coherent narratives that assist human understanding.
Analytical search, in contrast, targets well-defined analytical questions with the explicit goal of reaching a correct and verifiable conclusion. 
Rather than maximizing coverage, analytical search emphasizes sufficiency and relevance of evidence with respect to a specific analytical objective. 
The task structure is therefore markedly different. 
Deep research tasks are often loosely structured and evolve dynamically during execution, with stopping criteria that are implicit, heuristic, or user-driven. 
Analytical search tasks, by contrast, are explicitly structured, with clear analytical goals, decomposed sub-tasks, and well-defined termination conditions-such as satisfying evidential requirements or completing a reasoning chain.

Differences also emerge in how reasoning and evidence are handled. 
In deep research systems, reasoning is typically implicit and narrative-driven, embedded within long-form synthesis that prioritizes coherence and readability. 
Evidence is often aggregated descriptively rather than interrogated analytically. 
Analytical search instead requires explicit, step-wise reasoning tightly coupled with evidence selection, transformation, and validation. Each reasoning step is expected to serve a functional role in advancing the analysis, rather than contributing to a general narrative.
Accordingly, the evaluation focus diverges. 
Deep research systems are commonly assessed based on usefulness, coverage, and readability of the synthesized output. 
Analytical search systems, by contrast, are evaluated on the correctness of conclusions, the completeness of critical evidence, and the traceability of the reasoning process that leads from query to conclusion.

\subsubsection{Analytical Search vs. Agentic Databases} \hfill \break
Agentic databases and database agents~\cite{hu2025drama,sun2025agenticdata,tang2025llm} represent another related paradigm, emphasizing natural language interaction with structured data systems. 
These systems typically operate over well-defined, structured data and focus on translating user intent into executable queries (e.g., SQL), optimizing query execution, and returning precise results. 
Their strength lies in accurate query formulation, efficient execution, and deterministic correctness within a closed-world schema.
Analytical search extends beyond this scope in both data coverage and system objectives. 
Rather than operating solely on structured databases, analytical search must function over heterogeneous and evolving data environments, integrating structured records, semi-structured sources, unstructured text, and open web data, while optimizing indexes within the online workflow. 
It often requires combining quantitative results from databases with qualitative evidence from documents, reports, or news articles, exceeding the assumptions of most agentic database systems.

More fundamentally, analytical search reframes the role of search itself. 
Instead of viewing search as query execution, analytical search treats it as analysis orchestration. 
The system must coordinate retrieval, fusion, validation, and possibly iterative refinement across multiple data modalities and tools. 
Query execution is only one component in a broader analytical pipeline whose ultimate goal is not data access per se, but evidence-based problem solving.

\begin{figure*}[t]
    \centering
    \includegraphics[width=0.99\textwidth]{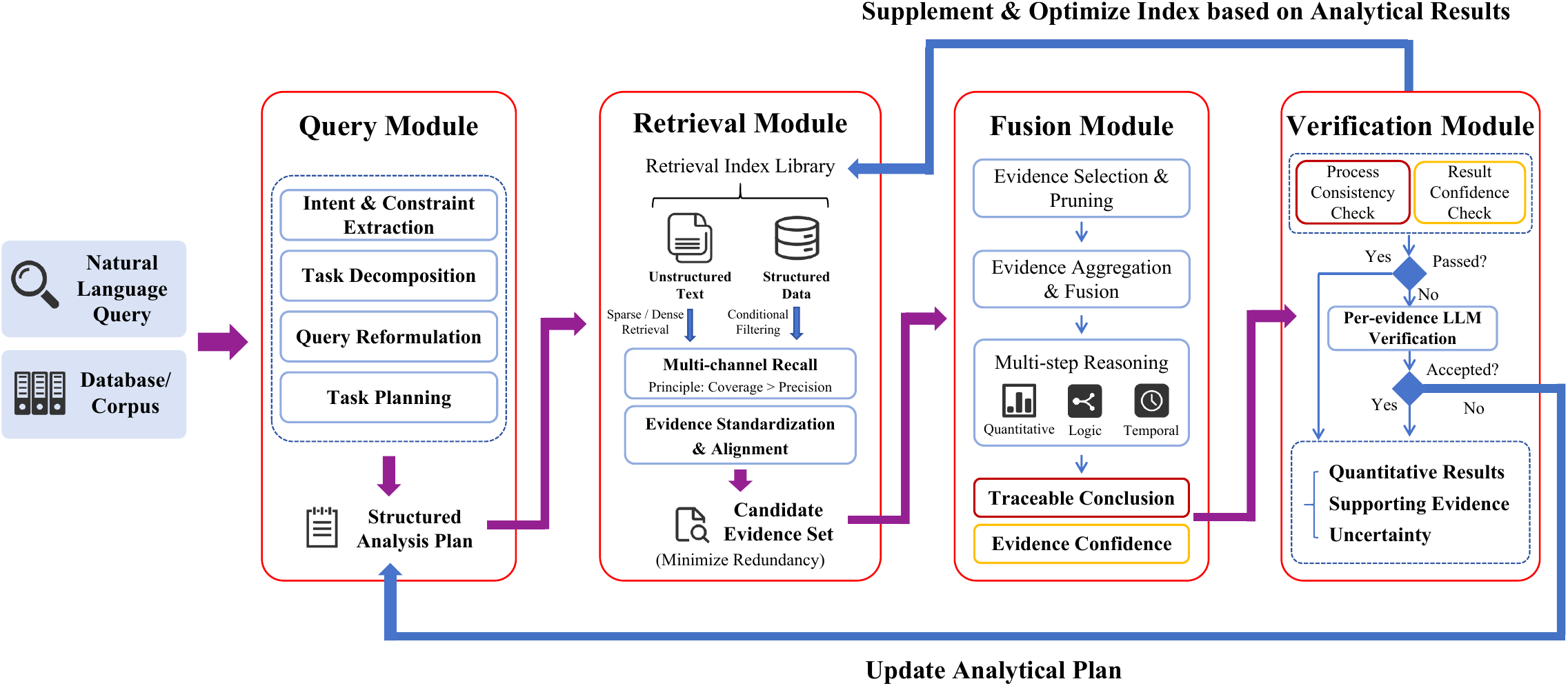}
    \caption{A conceptual framework for analytical search.}
    \label{fig:framework}
\end{figure*}

\section{Conceptual Framework} \label{sec:frame}
Based on the aforementioned characteristics and challenges, analytical search requires a system architecture that goes beyond traditional IR (+LLM) pipelines. 
Instead of treating search as a single-step mapping from query to answer, analytical search should support end-to-end analytical problem solving, in which intermediate task structures, evidence states, and reasoning steps are explicit and integral to system behavior.
At a high level, analytical search can be viewed as an analytical workflow that transforms a user’s natural language query into an explainable and verifiable conclusion. 
This workflow proceeds through a sequence of stages: \textit{analytical query → task modeling → evidence retrieval → fusion → verification → conclusion} (as shown in Figure~\ref{fig:framework}). 
Each stage serves a distinct module and produces explicit intermediate representations that can be inspected, revised, or validated.
By separating what needs to be solved, what evidence is required, how evidence is fused, and how conclusions are justified, this workflow enables transparency, controllability, and robustness when addressing complex analytical information needs.
In this section, we present a conceptual framework that operationalizes this workflow, comprising four core components: query modeling, evidence retrieval, reasoning-intensive fusion, and adaptive verification, which collectively form a coordinated system for evidence-grounded analytical problem-solving.
It distinguishes analytical search from conventional IR systems that largely collapse analysis into relevance ranking. 

\subsection{Query Module} \label{subsec:query_module}
The query module serves as the system’s entry point and is responsible for transforming complex, ambiguous, informal, or underspecified natural language queries into explicit analytical (sub-)task structures that the system can execute.
Additionally, analytical queries are often characterized by implicit intent, unspoken constraints, and non-expert terminology. 
Therefore, the query module must go beyond keyword interpretation or intent classification and instead construct a structured analytical plan that captures both what the user is asking and what must be done to answer it. 
Its core capabilities include:
\begin{itemize}[leftmargin=*]
    \item \textbf{Intent understanding and constraint extraction}, identifying analytical goals, target entities, temporal scopes, potential constraints, comparison baselines, and evaluation criteria, even when they are not explicitly stated.
    \item \textbf{Task decomposition}, breaking complex analytical queries into logically coherent sub-tasks that can be addressed independently and then recombined.
    \item \textbf{Query reformulation}, resolving ambiguity, normalizing colloquial or non-professional expressions, and making implicit assumptions explicit so that downstream retrieval, fusion, and reasoning are well-posed.
    \item \textbf{Retrieval and tool planning}, determining which data sources, retrieval paradigms, or analytical tools are appropriate for each sub-task and how they should be orchestrated.
\end{itemize}
The output of the query module is not simply rewritten queries, but an executable analytical plan that guides subsequent stages.

\subsection{Retrieval Module}
In analytical search, retrieval plays a role that is fundamentally different from that in traditional IR systems. 
Rather than functioning primarily as a mechanism for information access, the retrieval module is responsible for task-conditioned evidence acquisition, assembling the evidence necessary to support downstream fusion, analysis, and reasoning. 
This module warrants explicit separation because failures in analytical search are more often attributable to missing, biased, or insufficient evidence than to shortcomings in language generation. 
As a result, retrieval in analytical search prioritizes evidential completeness and analytical coverage over narrow relevance ranking. 
Unlike conventional retrieval pipelines that emphasize precision-oriented top-$k$ results and ranking sharpness, analytical search adopts a \textbf{recall}-oriented perspective, ensuring that all potentially relevant evidence needed for analysis is available downstream.
Moreover, retrieval is no longer confined to a single document type or data modality; instead, it operates over heterogeneous sources, encompassing unstructured text, semi-structured documents, tables, statistical data, and structured records, all of which may be jointly required to support a coherent analytical conclusion.
The retrieval module is also responsible for evidence normalization and alignment, that is, harmonizing representations across sources so that evidence can be meaningfully compared, aggregated, or reasoned over.
This module thus constructs an evidence space tailored to the analytical task, rather than merely returning a ranked list of documents.

\subsection{Fusion Module}
The fusion module constitutes the analytical core of the system, operating over the retrieved evidence to perform structured, multi-step inference that incrementally advances toward an analytical conclusion.
In analytical search, fusion is explicitly modeled as a decision process with traceable intermediate states, where the system autonomously selects among possible reasoning steps, including analytical operations and utilizing external tools, based on the evolving problem state and evidence context. 
These steps, such as choosing which evidence to focus on, deciding when and how to aggregate or filter information, or determining which analytical operation to apply, are not pre-fixed stages but flexible reasoning actions that are dynamically invoked as needed. 
Within this process, the fusion module is responsible for identifying and aggregating relevant evidence, executing logical, temporal, and quantitative operations such as cross-period comparison, causal inference, trend detection, and numerical computation through internal calculations or by explicitly calling external tools such as coding and SQL.
It also ensures that all intermediate inferences and conclusions remain firmly grounded in the available evidence. 
By making reasoning-based fusion process explicit, stateful, and inspectable, this module reduces the risk of hallucination and enables robust, explainable analytical problem solving, clearly distinguishing analytical search from purely generative approaches.

\subsection{Verification Module}
Verification is not an optional add-on in analytical search; it is an intrinsic system component. 
Because analytical conclusions often inform decisions or interpretations in high-stakes domains, they must be auditable and trustworthy.
The verification module provides analytical accountability by examining both the reasoning process and the resulting conclusions. Its core functions include:
\begin{itemize}[leftmargin=*]
    \item \textbf{Process-level consistency checking}, ensuring that reasoning steps are logically coherent and that evidence is exploited appropriately. 
    \item \textbf{Result-level validation}, confirming that conclusions are supported by sufficient and non-contradictory evidence.
    \item \textbf{Adaptive control}, triggering backtracking or additional retrieval when evidence is insufficient or conflicting, while allowing verification to be skipped or minimized when confidence is already high (e.g., when a query can be resolved by a simple, well-defined database operation).
\end{itemize}

Through this adaptive verification mechanism, analytical search balances rigor and efficiency, ensuring reliability without unnecessary computational overhead.

\subsection{An Illustrative End-to-End Example}
To illustrate how the proposed modules interact in practice, consider the query "\textit{Did the introduction of traffic cameras reduce pedestrian accidents in City X over the past five years?}"
The query module first reconstructs its implicit structure by identifying the intervention variable (i.e., "camera deployment"), the outcome variable (i.e., "pedestrian accident frequency"), the temporal scope, and the need for a before-and-after comparison with potential control factors. 
It decomposes the task into sub-steps, including determining the camera deployment timeline, retrieving structured accident statistics, collecting relevant environment variables, and other contextual information. 
The retrieval module then conducts recall-oriented, multi-path evidence acquisition. 
For example, it may obtain structured data through executable database queries (e.g., Text-to-SQL for yearly accident counts) and use sparse and dense retrieval methods to collect unstructured reports and policy documents. 
The fusion module then performs temporal alignment and quantitative aggregation, and invokes external analytical tools if needed (e.g., statistical testing or regression) to estimate whether accident rates significantly declined after camera deployment. 
Finally, the verification module checks the logical consistency, evidential sufficiency, and cross-source agreements on the evidence, triggering additional retrieval if necessary, and produces a traceable and confidence-calibrated conclusion.

\section{Potential Research Directions and Challenges} \label{sec:method}
Based on the conceptual framework introduced above, this section outlines key research directions for realizing analytical search in practice, together with the fundamental challenges that shape their realization.
We discuss a set of methodological perspectives and their accompanying challenges and believe that addressing these intertwined opportunities and constraints is central to building robust, scalable, and accountable analytical search systems.

\subsection{Reasoning as Sequential Decision Making}
From a methodological perspective, analytical search can be naturally formulated as a sequential decision-making problem, in which the system incrementally constructs an analytical solution through a series of interdependent reasoning actions. 
Rather than viewing query understanding, retrieval planning, evidence fusion, and verification as loosely coupled stages optimized in isolation, analytical search treats them as a coordinated reasoning process that spans the entire analytical workflow. 
At each step, the system must decide how to interpret and decompose the query, from which data source and in what manner the evidence is retrieved, how to calculate and fuse the current evidence, and whether the current conclusions require further validation or revision.
Within this formulation, analytical search can be viewed as an agent interacting with an environment composed of heterogeneous data sources, retrieval tools, intermediate evidence states, and evolving reasoning contexts. 
Actions correspond to analytical decisions such as query decomposition and reformulation, retrieval planning and control, evidence selection and aggregation, analytical inference, and adaptive triggering of verification. 
The state captures the current analytical context, including the task structure derived from the query, the evidence collected so far, intermediate reasoning results, and confidence estimates. 
This perspective explicitly models analytical reasoning as a controlled, multi-step process rather than an implicit by-product of text generation.

\subsubsection{Structural Design} \hfill \break
Under the sequential decision-making view, analytical search is realized through a coordinated chain of reasoning decisions distributed across three core modules: the \textit{query module}, which initiates and structures the analytical trajectory by interpreting intent and planning sub-tasks; the \textit{fusion module}, which drives the progression of analysis through stateful, multi-step reasoning over evidence; and the \textit{verification module}, which monitors, evaluates, and adaptively regulates the reasoning process to ensure correctness and evidential sufficiency.
The \textit{query module} serves as the entry point of analytical reasoning. 
Since analytical queries are often underspecified, ambiguous, or expressed in non-expert language, making direct retrieval ineffective, the query module has to perform active reasoning over the query itself, transforming a natural language request into an executable analytical plan, including identifying the underlying analytical intent, extracting implicit constraints and assumptions, decomposing the query into interdependent sub-tasks, and reformulating these sub-tasks into well-posed retrieval or analysis requests as discussed in \S\ref{subsec:query_module}.
For example, quantitative aggregation tasks may be routed to structured databases, while causal or contextual analysis may require unstructured text collections or external reports. 
Through this planning process, the query module directly shapes the downstream evidence space and constrains the reasoning paths available to the system.
The \textit{fusion module}, in turn, acts as the core reasoning agent that operates over the retrieved evidence to advance the analysis toward a conclusion. 
Rather than passively generating text conditioned on retrieved documents, the fusion module performs a sequence of explicit reasoning actions, such as selecting relevant evidence, aligning information across time or entities, performing filtering, comparisons, or aggregations, and drawing intermediate inferences. 
These actions are stateful and interdependent: each reasoning step updates the analytical context and informs subsequent decisions.
Complex analytical queries typically require chaining multiple such steps, making long-horizon planning and stability essential.
To ensure correctness and robustness, reasoning-based fusion in analytical search is tool-augmented by design. 
The fusion module selectively invokes external tools, such as SQL engines, code execution environments, or domain-specific analytical functions, when precise computation, aggregation, or simulation is required. 
Tool invocation is also treated as an explicit reasoning action rather than an opaque internal operation, allowing the system to offload deterministic computation while maintaining control over the overall analytical logic. 
This separation improves numerical accuracy, reproducibility, and interpretability.
Throughout this sequential reasoning process, the \textit{verification module} provides adaptive analytical control. 
Verification is not executed uniformly at every step; instead, it is triggered when confidence is low, evidence is conflicting, or conclusions carry high stakes. 
Verification actions may include consistency checking across evidence sources, validation of intermediate results, or backtracking to earlier stages to acquire additional evidence. 
By integrating verification into the reasoning loop, the system balances rigor and efficiency, avoiding both premature conclusions and unnecessary computation.

\subsubsection{Training Strategies and Challenges} \hfill \break
This sequential decision-making view enables end-to-end optimization of the analytical search workflow. 
Reinforcement learning (RL) and related optimization frameworks like GRPO~\cite{guo2025deepseek,shao2024deepseekmath} can be applied to jointly optimize the aforementioned reasoning actions of the three modules based on task-level rewards without an extra reward model. 
Such reward signals can also be derived from multiple complementary criteria.
For example, the most direct signal comes from the correctness of the final analytical conclusion, reflecting whether the system ultimately resolves the analytical task as intended. 
Beyond final outcomes, intermediate rewards can be defined based on evidence quality, encouraging the selection of sufficient, diverse, and non-contradictory evidence; reasoning stability, favoring coherent and logically consistent reasoning paths; and efficiency, penalizing unnecessary reasoning steps, excessive tool invocations, or redundant verification when confidence is already high.
As a result, analytical search systems can learn to adapt their reasoning strategies in an end-to-end manner across different analytical scenarios and metrics, determining when to explore broadly, when to reason deeply, and when to terminate with a confident, evidence-backed conclusion.

However, formulating analytical search as sequential decision making also exposes a set of fundamental training challenges that are intrinsic to this paradigm. 
Unlike conventional retrieval or generation tasks, as mentioned above, analytical reasoning policies are learned under highly underspecified intent.
As a result, training signals must encourage the system to actively hypothesize, test, and revise latent analytical intent through interaction with evidence, rather than merely imitate fixed query–answer mappings. 
This creates a tension between flexibility and overcommitment: models must reason beyond what is explicitly stated, while avoiding spurious assumptions that lead the reasoning process astray. 
Designing learning objectives and algorithms that support robust intent inference in such open-ended settings remains a central challenge for sequential reasoning-based analytical search.

A closely related difficulty lies in the stability of long-horizon reasoning. 
The analytical workflow often requires extended decision sequences that span query decomposition, multi-round retrieval, tool invocation, evidence fusion, and verification. 
Errors introduced early in the trajectory, such as incorrect task decomposition or biased retrieval decisions, can propagate and compound, ultimately invalidating the conclusion. 
From a sequential decision-making standpoint, this corresponds to the problem of maintaining policy consistency and recoverability over long horizons in a partially observable and non-stationary environment. 
Therefore, an effective training strategy should equip the agent with the ability to detect and correct intermediate mistakes, revise earlier decisions when new evidence emerges, and balance exploration and exploitation across reasoning steps. 
Purely myopic or step-local optimization is insufficient; instead, training must explicitly account for the cumulative effects of early decisions on downstream outcomes.

Finally, it poses a distinctive challenge in learning from sparse and delayed rewards. 
In many analytical tasks, meaningful supervision is only available at the end of the reasoning process, when the correctness and defensibility of the ultimate conclusion can be assessed. 
Intermediate reasoning steps, such as evidence selection, aggregation choices, or verification triggers, often lack explicit ground-truth labels, making credit assignment inherently difficult. 
Within the sequential decision-making framework, this necessitates the design of reward structures and auxiliary learning signals that align intermediate actions with long-term analytical objectives. 
Potential approaches include incorporating evidence-quality rewards, reasoning-consistency constraints, or efficiency-aware penalties that discourage unnecessary retrieval, tool usage, or verification when confidence is already high. 
Integrating such signals allows analytical search systems to learn not only what conclusions to reach, but how to reason efficiently and reliably toward them, reinforcing the view that training is inseparable from the system’s sequential reasoning formulation.

\subsection{Recall-Oriented Multi-Path Retrieval}
A core methodological departure of analytical search from traditional IR systems lies in its recall-first orientation. 
Classical IR systems are typically optimized for precision at top-$k$, under the assumption that users can manually inspect a small set of highly ranked results. 
Analytical search, however, shifts the burden of analysis from the user to the system. As a consequence, missing critical evidence is often more damaging than introducing moderate amounts of noise. 
Analytical failures are therefore more likely to arise from insufficient or biased evidence coverage than from imperfect ranking. 
This makes \textit{recall}, rather than precision, the primary optimization objective in the retrieval stage of analytical search.
Besides, from an analytical perspective, relevance is not a binary or surface-level notion. 
A document or data record that appears only weakly relevant in isolation may become indispensable when combined with other evidence in a reasoning chain. 
Analytical queries frequently require triangulating facts across time, sources, or modalities, and such triangulation is only possible if the relevant evidence is present in the candidate pool. 
Consequently, analytical search systems must prioritize evidential completeness, even at the cost of admitting additional noise that can later be filtered through reasoning and verification.

\subsubsection{Retrieval Routing} \hfill \break
To support this objective, analytical search adopts a multi-path retrieval strategy, in which different retrieval mechanisms are invoked depending on the structure of the query and the nature of the underlying data. 
Rather than relying on a single retrieval paradigm, the system dynamically routes sub-tasks through specialized retrieval paths that are best suited to their evidential requirements.
For \textit{structured data}, analytical search emphasizes precise, executable retrieval. 
When the analytical task involves counts, aggregations, or well-defined constraints, such as incident statistics or financial indicators, the system first transforms natural language sub-queries into structured representations through a Text-to-SQL~\cite{qin2022survey,luo2025natural,gao2023text,liu2025survey} model. 
These structured queries are then executed directly against databases, enabling accurate retrieval of quantitative evidence with well-defined semantics. 
This pathway ensures determinism and correctness for analytically critical operations such as filtering, grouping, and aggregation, which are difficult to approximate reliably through text-based retrieval alone.
On the other hand, for \textit{unstructured data}, analytical search employs sparse and dense retrieval methods to construct a broad candidate evidence set. 
Sparse retrieval~\cite{robertson2009probabilistic,ponte2017language} is effective for capturing exact matches, domain-specific terminology, and named entities, while dense retrieval~\cite{karpukhin2020dense,zhan2020repbert} provides semantic generalization across paraphrases and contextual variations.
Given the recall-first objective, these methods are used in a complementary manner to maximize coverage. 
However, since high recall inevitably introduces noise, the system can incorporate a subsequent re-ranking~\cite{pradeep2023rankvicuna,nogueira2019multi} stage that evaluates candidates not merely by topical relevance, but by their reasoning value—that is, their potential contribution to analytical reasoning, such as supporting comparisons, temporal alignment, or causal inference in the fusion module.
Together, these structured and unstructured retrieval paths form a unified, recall-oriented retrieval framework tailored to analytical search. 
By explicitly embracing recall as the dominant objective and routing queries through modality-appropriate retrieval strategies, it establishes a robust evidential foundation upon which reliable reasoning and verifiable conclusions can be built.

\subsubsection{Challenges} \hfill \break
Despite its conceptual advantages, this retrieval approach also introduces a set of non-trivial challenges. 
A primary difficulty lies in the trade-off between evidential coverage and downstream efficiency. 
While maximizing recall is essential to avoid missing analytically indispensable evidence, excessively low precision can severely burden subsequent fusion and filtering stages. 
A noisy or redundant evidence pool increases the difficulty of reasoning-based fusion, amplifies the risk of spurious correlations, and leads to unnecessary tool invocations and verification overhead. 
As a result, recall-oriented retrieval cannot be treated as indiscriminate expansion; it must still maintain a minimal level of analytical precision to ensure that downstream reasoning remains tractable and stable.
Beyond this precision–recall tension, multi-path retrieval introduces additional coordination challenges: different retrieval paths often return evidence with heterogeneous semantics, granularity, and reliability. 
Aligning these results into a coherent evidence space, resolving conflicts between sources, and avoiding systematic bias toward any single retrieval path remain non-trivial problems. 
Moreover, analytical queries may dynamically shift retrieval priorities as reasoning progresses, requiring retrieval policies that can adaptively rebalance paths rather than execute them in a fixed, one-shot manner. 
Addressing these challenges is crucial for ensuring that recall-oriented multi-path retrieval serves as an enabler, rather than a bottleneck, for efficient and reliable analytical reasoning.

\subsection{Dynamic and Task-Aware Index Organization}
Analytical search operates in settings where information needs are open-ended, evolving, and often difficult to anticipate in advance. 
As a result, the underlying index cannot be treated as a static structure constructed once and queried indefinitely. 
Instead, analytical search requires a dynamic and task-aware index organization in which the index itself evolves in response to real analytical workloads. 
In this paradigm, the index embodies long-term analytical priors, capturing accumulated knowledge about what kinds of information and structures are repeatedly useful for analysis, while retrieval mechanisms handle short-term task execution, selecting evidence relevant to a specific query or sub-task.
This distinction highlights a fundamental shift from traditional IR systems, where the index is largely fixed and query-independent. 
In analytical search, the index functions as an evolving analytical substrate that is continuously refined through interaction. 
Index evolution is driven by observed query patterns, reasoning trajectories, and evidence usage, allowing the system to progressively internalize structural regularities that are difficult to define a priori.

For structured data, this need is particularly evident. 
In many real-world databases, it is impractical to predefine all possible columns or analytical dimensions that users may require. 
Analytical queries frequently introduce new perspectives, such as derived metrics, composite attributes, or domain-specific categorizations, that are absent from the original schema. 
Therefore, the analytical search system can benefit from query-driven index evolution, in which user queries and analytical plans trigger the discovery and construction of new columns or views in the background. 
These dynamically induced structures allow future queries to be answered more directly and reliably, effectively transforming latent analytical concepts into explicit, indexed representations.
A similar dynamic also applies to unstructured text collections. 
Not all documents contribute equally to analytical reasoning, and their importance often becomes apparent only through repeated use in analytical workflows. 
By observing interaction histories, such as which documents are frequently retrieved, selected during reasoning, or cited in validated conclusions, analytical search systems can expand and reorganize textual indexes to emphasize high-value content. 
Core, representative, or frequently referenced documents can be enriched with additional index entries, semantic annotations, or alternative representations, making them more readily accessible to retrieval components in subsequent analyses.

Collectively, these mechanisms redefine the index as more than a passive data structure. 
In analytical search, the index becomes a living representation of accumulated analytical experience, shaped by real tasks rather than predefined assumptions. 
By enabling query-driven index evolution, analytical search systems establish a feedback loop in which past analytical activity informs future efficiency, robustness, and coverage—laying the groundwork for scalable, long-term analytical intelligence.
However, on the other hand, such an indexing strategy also introduces practical challenges for analytical search. 
Aggressively evolving indexes in response to observed analytical workloads can improve efficiency, but may also risk overfitting to historical query distributions, leading to brittle behavior when analytical intents shift. 
Moreover, enriching indexes with derived structures, annotations, or analytical views increases storage and maintenance costs, and may blur the boundary between indexing and reasoning if not carefully controlled. 
Therefore, balancing adaptability with stability, ensuring that index evolution captures durable analytical value rather than transient patterns, remains a key challenge for deploying analytical search at scale.

\subsection{Evaluation Principles}
Evaluating analytical search systems requires moving beyond traditional relevance-based metrics that dominate classical IR evaluation. 
Because analytical search is designed to solve complex analytical problems rather than merely retrieve topically relevant documents, its evaluation must reflect analytical contribution, reasoning quality, and system efficiency. 
This necessitates a multidimensional evaluation framework that captures both outcomes and processes.
Here, we provide five important evaluation dimensions for consideration:
\begin{itemize}[leftmargin=*]
    \item \textbf{Correctness of the Conclusion}.
    The primary evaluation criterion in analytical search is whether the system reaches a correct and reasonable conclusion. 
    Unlike factoid question answering, conclusions may involve aggregation, comparison, or inference, and correctness must be assessed with respect to task-specific ground truth, expert judgment, or validated analytical outcomes.
    \item \textbf{Recall of Critical Evidence}.
    Analytical search systems must be evaluated on their ability to retrieve and utilize critical evidence required for sound reasoning. 
    Missing key evidence can invalidate an otherwise plausible conclusion. 
    Evaluation should therefore emphasize evidential recall and coverage, particularly for evidence that plays an essential role in the reasoning chain.
    \item \textbf{Logical Consistency}.
    Beyond the final conclusion, the internal reasoning process should be logically coherent.
    Intermediate steps should follow valid inferential patterns, avoid contradictions, and maintain consistency across evidence sources and analytical assumptions. 
    Logical inconsistency, even when leading to a correct answer by chance, should be penalized.
    \item \textbf{Traceability and Explainability}.
    Analytical conclusions should be traceable back to explicit evidence and reasoning steps. 
    Evaluation should assess whether the system can provide transparent explanations that expose how evidence was selected, transformed, and combined to support the final conclusion. 
    This criterion is especially important for trust, auditability, and error analysis.
    \item \textbf{Efficiency}.
    Finally, analytical search systems must be evaluated on efficiency, including the number of reasoning steps, tool invocations, retrieval rounds, and verification actions required to reach the conclusion. 
    Efficient analytical reasoning reflects not only computational performance but also the system’s ability to adaptively control depth and avoid unnecessary operations.
\end{itemize}

While these principles outline what should be evaluated, a core difficulty is the absence of a single gold-standard reasoning path. 
For many analytical tasks, multiple reasoning trajectories differing in decomposition strategies, evidence ordering, or analytical operations may all lead to defensible conclusions. 
Consequently, evaluation frameworks that assume a fixed reference process or a unique correct output are ill-suited for analytical search. 
Systems must be assessed in a manner that tolerates procedural diversity while still enforcing constraints on evidential sufficiency, logical coherence, and analytical validity.
Another inherent challenge arises from the multiplicity and conditionality of valid outcomes. 
Since analytical conclusions are often contingent on assumptions, scope choices, or evidential framing that may not be uniquely determined by the query alone. Different, yet reasonable, analytical interpretations can therefore yield different conclusions without any being strictly incorrect. 
This characteristic undermines evaluation protocols based purely on answer matching or surface-level correctness. 
Therefore, evaluation must shift toward assessing whether a conclusion is justified given the evidence and assumptions made, and whether those assumptions are explicitly surfaced and internally consistent, further reinforcing the importance of evidence-level and process-level evaluation, rather than treating the final answer as an isolated artifact.
In addition, human-in-the-loop evaluation incorporating expert judgment may be unavoidable for complex analytical tasks, especially in high-stakes domains where correctness, accountability, and interpretability outweigh scalability concerns. 
Together, these considerations position evaluation not merely as a benchmarking exercise but as a significant component of analytical search system design, shaping how systems reason, justify conclusions, and expose uncertainty.
\section{Conclusions} \label{sec:conclu}
In this paper, we introduce a novel search paradigm, \textbf{analytical search}, which represents a distinct and increasingly important information-seeking paradigm that cannot be adequately addressed by traditional relevance-oriented IR systems or by straightforward combinations of retrieval and generation. 
By formalizing analytical information needs, articulating a new analytical search paradigm, and presenting a conceptual system framework together with research directions and challenges, we highlight how analytical search shifts the focus from naive information finding to evidence-grounded problem solving, from answer fluency to conclusion correctness, and from static retrieval to reasoning-aware analytical workflows. 
More broadly, analytical search serves as a unifying framework across information retrieval, natural language processing, and database systems, bringing together retrieval, reasoning, structured querying, and verification under a common analytical objective. 
As complex analytical information needs continue to proliferate across various domains, addressing analytical queries and building analytical search systems are both foundational challenges and critical opportunities for the IR community. 
We therefore call on the community to recognize analytical search as a first-class research problem and to invest in its systematic study.

\begin{acks}
This work is supported by the Research Project of Quan Cheng Laboratory, China (Grant No. QCL20250105)
\end{acks}

\bibliographystyle{ACM-Reference-Format}
\bibliography{sample-base}

\end{document}